\documentclass{article}
\usepackage{ijcai22}

\usepackage{times}

\usepackage{soul}
\usepackage{url}
\usepackage[hidelinks]{hyperref}
\usepackage[utf8]{inputenc}
\usepackage[small]{caption}
\usepackage{graphicx}
\usepackage{amsmath}
\usepackage{booktabs}
\usepackage[dvipsnames]{xcolor}
\usepackage{float}

\newcommand{\anonymous}[2]{#2}

\newcommand{\outline}[1]{}
\newcommand{\continuingissues}[1]{}

\newcommand{\KnowledgeBase}[0]{Knowledge Base}

\newcommand{\TeachingInterface}[0]{Teaching Interface}
\newcommand{\knowledgebase}[0]{\textsl{knowledge base}}
\newcommand{\machinelearner}[0]{\textsl{machine learner}}
\newcommand{\teachinginterface}[0]{\textsl{teaching interface}}

\newcommand{\UXUI}[0]{UX/UI}
\newcommand{\UX}[0]{UX}
\newcommand{\UI}[0]{UI}
\newcommand{\ML}[0]{ML}

\newcommand{\MT}[0]{MT}

\title{Human-AI Interaction Design in Machine Teaching}

\anonymous{
\author{
\affiliations
\textbf{Content Areas:} Intelligent User Interfaces, 
Knowledge Aided Learning, \\
Dialogue and Interactive Systems,
NLP Tools,
Cognitive Systems \\}
} {
\author{
Karan Taneja\footnote{Corresponding author}\and
Harshvardhan Sikka\And
Ashok Goel\\
\affiliations
Georgia Institute of Technology
\emails
\{karan.taneja, ashok.goel\}@cc.gatech.edu, harshsikka@gatech.edu
}
}

\begin{document}


\maketitle

\begin{abstract}

Machine Teaching (\MT) is an interactive process where a human and a machine interact with the goal of training a machine learning model (\ML) for a specified task. 
The human teacher communicates their task expertise and the machine student gathers the required data and knowledge to produce an \ML~model. 
\MT~systems are developed to jointly minimize the time spent on teaching and the learner's error rate.
The design of human-AI interaction in an \MT~system not only impacts the teaching efficiency, but also indirectly influences the \ML~performance by affecting the teaching quality.
In this paper, we build upon our previous work where we proposed an \MT~framework with three components, viz., the \teachinginterface, the \machinelearner, and the \knowledgebase, 
and focus on the human-AI interaction design involved in realizing the \teachinginterface.
We outline design decisions that need to be addressed in developing an MT system beginning from an ML task.  
The paper follows the Socratic method entailing a dialogue between a curious student and a wise teacher.
\end{abstract}

\anonymous{}{
\vspace{-9pt}
}

\section{Introduction}

\textit{What is Machine Teaching (\MT)?}
\MT~is the process of training a machine learning (\ML) model through an interaction between a machine student and a human teacher. %
\MT~systems enables a wider community, beyond ML experts, to teach concepts to machine learners. 

\textit{Why is MT important?}
While ML research is devoted to improving performance of learning algorithms on various domains and tasks, MT research focuses on making development of ML models more accessible.
According to \cite{Simard2017MachineSystems}, intuitive, efficient, and friendly MT interfaces should be able to \textit{decouple} \MT~and \ML~processes to ensure that teachers do not require knowledge of the underlying ML algorithms, but only the task expertise.  
Further, \MT~research jointly aims to reduce the cost of creating ML models along with an increase in the performance. 
Previous research, such as work by \cite{Zhu2018AnTeaching,Liu2017IterativeTeaching,Zhu2015MachineEducation}, theoretically studied \MT~as an optimization problem described below where $\mathcal{D}$ is the dataset used for teaching, $\theta$ represents \ML~model parameters and $\eta$ is a scaling parameter.
\begin{equation*}
\begin{aligned}
    \min_{\mathcal{D},\theta} \quad & \text{TeachingRisk} (\theta) + \eta \cdot \text{TeachingCost}(\mathcal{D})\\
%
    \text{s.t.} \quad & \theta = \text{MachineLearning}(\mathcal{\mathcal{D}})\\
\end{aligned}
\end{equation*}
TeachingRisk measures the learner's error, which can be defined using a test set, with model parameters $\theta$ and TeachingCost measures the resources spent on teaching such as number of examples or teaching time.

\begin{figure}[ht!]
    \centering
    \includegraphics[width=\linewidth]{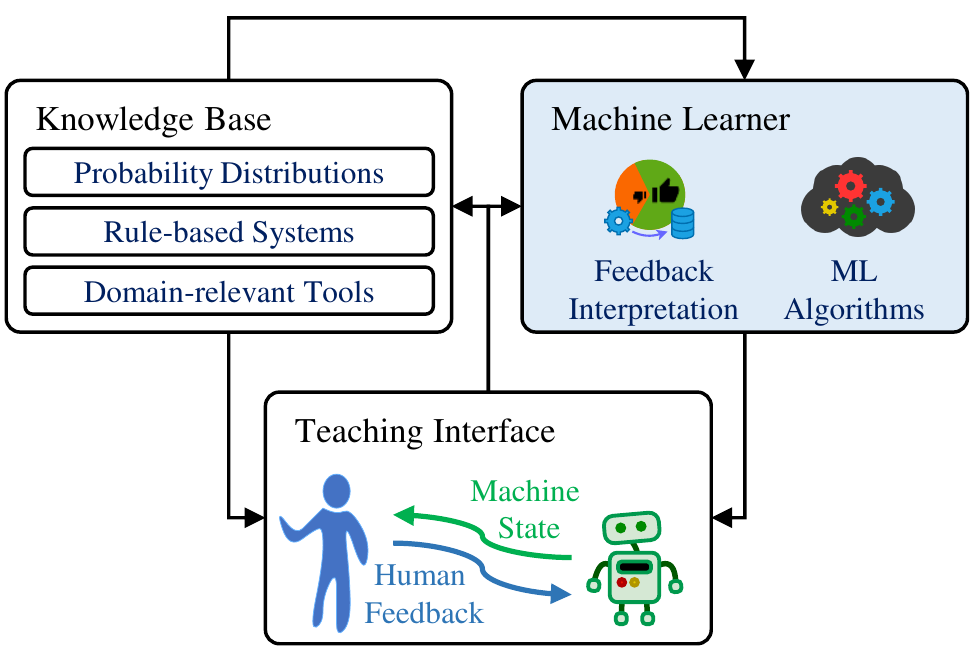}
    \caption{
    \textbf{Three main components of an \MT~system: \teachinginterface, \knowledgebase, and \machinelearner~\protect\cite{Taneja2022ATeaching}.} 
    The teaching feedback received at the \teachinginterface~is used by feedback interpretation mechanism and to build components of the \knowledgebase.
    The \machinelearner~feeds back into the \teachinginterface~by revealing its state, for instance, by providing confusing examples for teaching and supporting interpretable ML methods.
    The \knowledgebase~can augment teaching feedback to support feedback interpretation as well as provide assistance to teachers at the \teachinginterface. 
    }
    \label{fig:framework}
\end{figure}

\begin{figure*}[ht!]
    \centering
    \includegraphics[width=\linewidth]{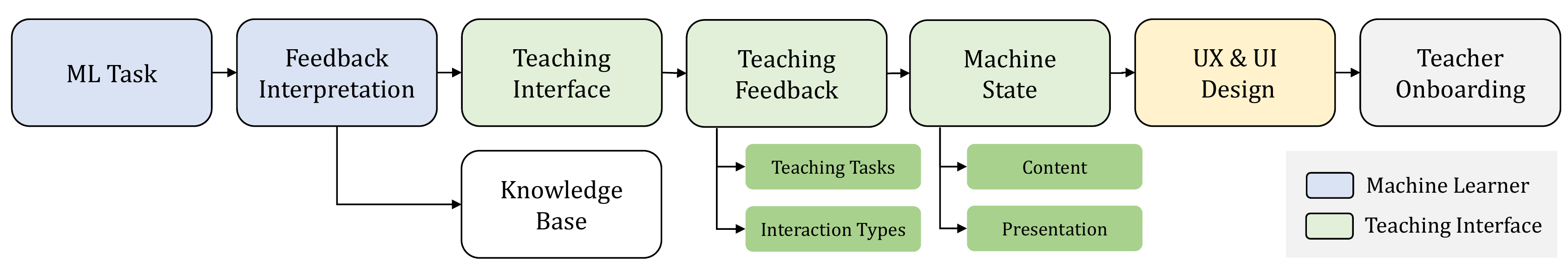}
    \caption{
    \textbf{Stages of MT system design process.}
    Once feedback interpretation mechanism and knowledge base components are known, we can decide the teaching tasks that human teachers must perform to support the feedback interpretation mechanism. 
    This is followed by specifying how machine state will be revealed to teachers for each teaching task. 
    The final steps include \UXUI~design and creating a process for onboarding new teachers.
    }
    \label{fig:design_flow}
\end{figure*}

\textit{What will we discuss in this paper?} 
In our previous work \cite{Taneja2022ATeaching}, we proposed a framework for designing \MT~systems by describing three components, viz., the \teachinginterface, the \machinelearner, and the \knowledgebase~(see Figure \ref{fig:framework}). 
We also introduced and experimented with our implementation of an MT system for text classification.
In this paper, we focus on various aspects of human-AI interaction design involved in realizing the \teachinginterface~for any \MT~system.
Figure \ref{fig:design_flow} traces the different stages of design process of an MT system. 
We will start with a brief discussion about \MT~systems and, then, examine communication at the \teachinginterface~from both the teacher and the machine point of view. 
We will also make comments on the user experience and user interface (\UXUI) in \MT~systems followed by a discussion about onboarding new teachers to use the \teachinginterface.  
At each stage, we outline design decisions to be addressed by developers of MT systems and provide instructive examples for additional insights.

\textit{What are the main contributions of this paper?} 
%
(i) We outline a design process for MT systems from the perspective of human-AI interaction (Section \ref{sec:human_ai_interaction}).
(ii) We discuss existing literature as examples in the above context.
(iii) We discuss research directions surrounding human-AI interaction in MT systems (Section \ref{sec:future_research}).

\section{Human-AI Interaction in Machine Teaching}
\label{sec:human_ai_interaction}

\textit{How does this work relate to the MT framework proposed in \cite{Taneja2022ATeaching}?} 
In our previous work, we introduced an MT framework with the \teachinginterface~component as the hub of human-AI interaction (Section \ref{subsec:MT_components}). 
This work focuses on two channels of communication noted in our previous work: 
the human-to-AI channel described as Teaching Feedback in Section \ref{subsec:teaching_feedback} and 
the AI-to-human channel described as Machine State in Section \ref{subsec:machine_state}.
We also briefly discuss the \UXUI~design in Section \ref{subsec:UI_UX} 
and the process of onboarding new teachers to use the \teachinginterface~in Section \ref{subsec:onboarding}.  
Throughout the paper, we will also discuss the MT system for text classification proposed in our previous work and introduce a web interface for the same system as an illustrative example.

\subsection{Components of MT Systems}
\label{subsec:MT_components}


\textit{What are the main components of an MT system?} MT systems have three components: \teachinginterface, \machinelearner, and \knowledgebase~\cite{Taneja2022ATeaching}.  The human teacher interacts with the \machinelearner~using the \teachinginterface~to train ML algorithms, and this process is supported by existing domain and task-specific knowledge present in the \knowledgebase. These components are outlined in Figure \ref{fig:framework}.

\subsubsection{Machine Learner}

\textit{What is feedback interpretation in the \machinelearner?} 
The first component of the \machinelearner~i.e. the \textit{feedback interpretation mechanism} describes how the teaching feedback will be used to train a machine learning model.
For example, in the MT system proposed in \cite{Taneja2022ATeaching} for text classification, data augmentation was used for feedback interpretation. 
In other words, the human feedback collected by the machine was used in a data augmentation process and this augmented data was used to train the ML model. 

\textit{What other strategies can be used for feedback interpretation?}
Different strategies broadly try to manipulate one or more parts of the training process. 
For instance, input features can be constructed from human feedback \cite{Godbole2004DocumentLabels,Settles2011ClosingInstances,Jandot2016InteractiveClassification}, 
the feedback may be used to penalize intermediate representations like attention maps \cite{Qiao2017ExploringAnswering}, 
training data can be augmented \cite{Taneja2022ATeaching}, 
or the loss function may be modified \cite{He2016Human-in-the-LoopParsing,Stiennon2020LearningFeedback,Kreutzer2018CanFeedback}.

\textit{Are ML algorithms only used for the final predictions?} 
No, ML models in MT systems are not only used for the specific task, but also to aid the teaching process by providing its state to the human teacher for interpretation, which is discussed later in Section \ref{subsec:machine_state}. 
For example, \cite{Taneja2022ATeaching} used ML model to calculate deemed importance of each word by the machine as shown in Figure \ref{fig:interaction_example}. 
\cite{Godbole2004DocumentLabels,Settles2011ClosingInstances} used ML model to find most influential terms for each class to assist in feature engineering. 
\cite{Godbole2004DocumentLabels,Settles2011ClosingInstances,Simard2014ICE:Problems,Taneja2022ATeaching} have all used ML models to suggest the most confusing examples for teaching using active learning.
\cite{He2016Human-in-the-LoopParsing} used output uncertainty of the sentence parser to create questions that can be asked to non-experts for feedback.
\cite{Ramos2020InteractiveModels} used their model to display information extraction predictions, allowing teachers to examine them and debug errors that they find.

\subsubsection{\KnowledgeBase}

\textit{What is a knowledge base?} 
Knowledge base~is an umbrella-term for existing or assembled resources that can be used to aid the machine teaching process, 
including probabilistic or generative models, rule-based systems, and software tools.

\textit{Where do we use knowledge?}
The \knowledgebase~is used for feedback interpretation by the \machinelearner, and for assisting human teachers through the \teachinginterface~to increase teaching efficiency.
As an example, \cite{Taneja2022ATeaching} used off-the-shelf BERT masked-language models to recommend word replacements to human teachers as shown in Figure \ref{fig:interaction_example}, and for data augmentation by the machine learner. 
Further, human feedback was assembled to construct a domain-specific dictionary and used to improve recommendations for word replacements. 
\cite{Godbole2004DocumentLabels,Settles2011ClosingInstances} used the teacher's feedback to associate and store \textit{evidence words} with each class, which are later interpreted as features for training the ML model.
\cite{Simard2014ICE:Problems} similarly allowed teachers to create and edit features defined by dictionaries.
\cite{He2016Human-in-the-LoopParsing} used teacher's feedback for parsing task to store relations between entities in sentence.
\cite{Ramos2020InteractiveModels} allowed teachers to create and edit entity schemas that are used for information extraction. 

\subsubsection{\TeachingInterface} \label{subsec:interface}

\textit{What happens at the teaching interface?} 
The \teachinginterface~is where human teacher interacts with the machine (a.k.a. the AI) with the goal of producing a trained ML model. 
%
%
Broadly speaking, the machine communicates its \textit{machine state} to human teacher, and the teacher responds back with \textit{teaching feedback} for the machine.

\textit{How does the human-AI communication work here?}  
The \textit{teaching feedback} represents the human-to-AI communication and \textit{machine state} characterizes the AI-to-human communication. 
The next two subsections discuss these two channels in great detail.

\subsection{Teaching Feedback}
\label{subsec:teaching_feedback}

\textit{Why is the role of teaching feedback in an \MT~system?}
%
As mentioned earlier, the teaching feedback is used by feedback interpretation mechanism to train the ML model. 
Also, components in \knowledgebase~may use teaching feedback to build and store knowledge.  

\textit{How is the teaching feedback presented to the machine?}
The teaching feedback is communicated by performing \textit{teaching tasks} involving different interaction types such as demonstrating, categorizing, sorting or evaluating  \cite{Cui2021UnderstandingLearning}.
Similar interaction types have also been used for developing cognitive systems capable of learning from human teachers in the \textit{Natural Training Interactions} framework proposed by \cite{Harpstead2018TowardsFramework}.

\begin{figure}[t!]
    \centering
    \includegraphics[width=\linewidth]{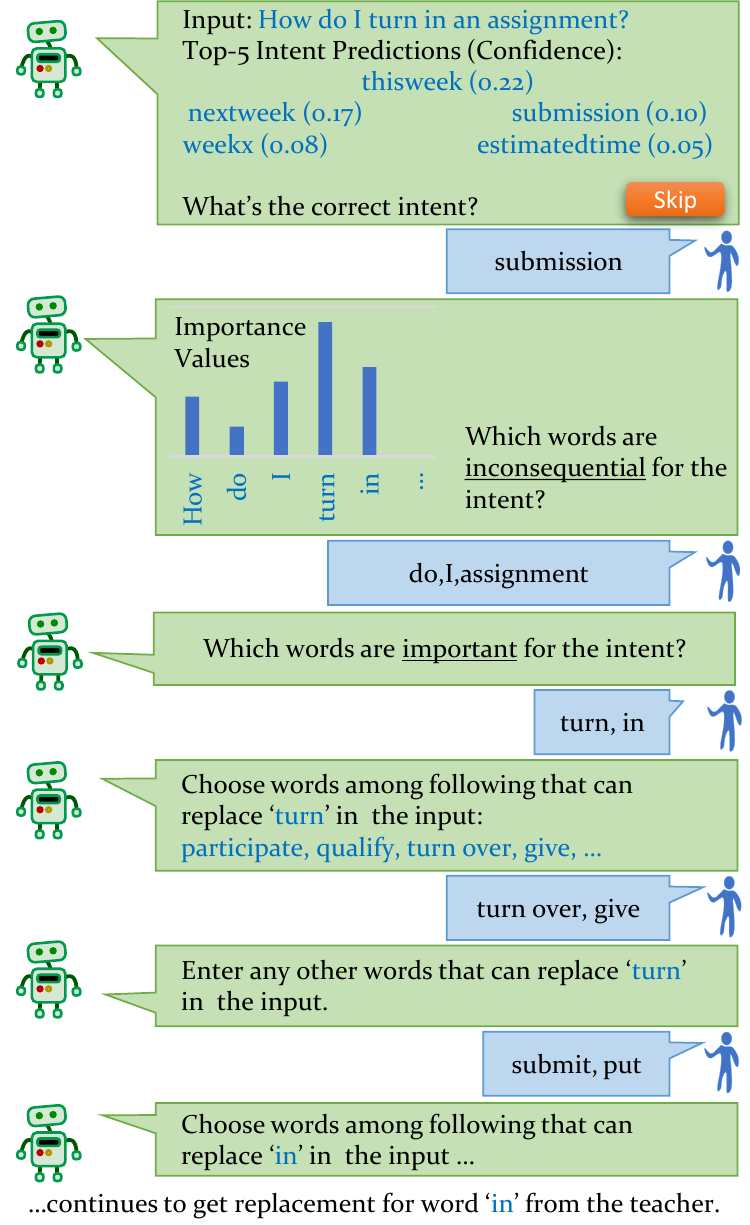}
    \caption{
    \textbf{An illustrative conversation between a human teacher and a machine student from \protect\cite{Taneja2022ATeaching}.}
    In this example, the human is training the machine on the intent classification task for questions asked to Jill Watson, a virtual teaching assistant in online classrooms \protect\cite{Goel2019JillEducation}.
    }
    \label{fig:interaction_example}
\end{figure}

\textit{How do we design these teaching tasks?} 
The teaching tasks are used to perform two main activities \cite{Ramos2020InteractiveModels}: (i) planning the teaching curriculum, and (ii) explaining knowledge related to subject domain.
The former includes tasks like example selection and bulk labeling.
For the latter, the teaching tasks depend on the feedback interpretation mechanism used by the machine learner. 
For example, MT system introduced in \cite{Taneja2022ATeaching}, used data augmentation as feedback interpretation strategy and, 
therefore, instructed human teachers to mark important and inconsequential words in a sentence classification task and asked teachers to validate replacements of important words (see Figure \ref{fig:interaction_example}).
\cite{Godbole2004DocumentLabels,Settles2011ClosingInstances} asked teachers to categorize influential words as a teaching task for using them during feedback interpretation to construct word-level features.
\cite{Simard2014ICE:Problems} also used features obtained by asking teachers to add relevant built-in features, or defining new ones, or using features that are learned model themselves.
\cite{He2016Human-in-the-LoopParsing} asked teachers to evaluate possible answers to questions about the relationships between entities for the sentence parsing task.
\cite{Ramos2020InteractiveModels} asked teachers to select examples, inspect and correct model predictions, create and edit entity schemas, and construct features.

\textit{How does the interaction design affect teaching feedback?}
Well-designed teaching tasks can improve teaching quality. 
The human teacher should be able to 
(i) understand the teaching tasks without ML expertise \cite{Simard2017MachineSystems}, and 
(ii) unambiguously determine the feedback response for the machine \cite{Cui2021UnderstandingLearning}.
In other words, complex and hard-to-understand tasks can be detrimental to teaching quality. 
Also, poor design of a teaching task, such asking teacher to select an option among several bad choices, limits the teacher in communicating their expertise \cite{Cui2021UnderstandingLearning}. 
Recollecting the two main goals of MT systems, we also note here that interaction design not only affects the teaching efficiency, but also the learner's performance \textit{because} of its influence on the quality of teaching feedback.

\textit{How can we serve teachers in efficiently providing feedback?}
The \UXUI~design for is one factor that determines teacher's efficiency. 
But further, to reduce redundancy, teachers should be able to concentrate on replacing learner's misconceptions i.e. on correcting learner's mistakes. 
%
But to realize the mistakes made by the machine, the machine must present its own interpretation of teaching examples which can then be corrected by the teacher.
This is the main role of \textit{machine state} and our next topic of discussion.

\subsection{Machine State for Human Teachers}
\label{subsec:machine_state}

\textit{What do you mean by machine state?}
Machine state describes the information presented to the human teacher to build a mental model of the current state of the machine.
In other words, machine state aims to help the teacher in understanding the progress and errors made by the learner with respect to the given task.
In making design decisions about the machine state revealed to the human teachers, two primary questions need to be answered \cite{Eiband2018BringingPractice}:
(i) what information about the machine is conveyed i.e. the \textit{content}, and 
(ii) how the information is conveyed i.e. the \textit{presentation}.  

\textit{What are some ways in which machine state can help the teaching tasks?} 
Several prior works include interfaces that communicate signal from the machine to assist the teacher in accomplishing the teaching feedback tasks. 
For example, \cite{Godbole2004DocumentLabels} displayed clusters of closely related confusing examples for bulk labeling.
Similar to \cite{Settles2011ClosingInstances}, they also presented influential words found using term-level active learning based on the current model to teachers for categorizing them into classes.
In addition to active learning, \cite{Simard2014ICE:Problems} computed and displayed various performance metrics in a graphical format and highlighted errors on training and test set to debug the ML model.
\cite{He2016Human-in-the-LoopParsing} communicated model uncertainty by presenting teachers with questions and possible answers which are constructed using output uncertainty of the parser.

\textit{How do we decide the content of the machine state?}
The content of the machine should be relevant and useful for providing feedback in the teaching tasks.
The teacher should be able to 
(i) understand current limitations of the machine by perceiving the machine state 
and, then,
(ii) correct mistakes by performing the teaching task. 
As a rule of thumb, 
interpretations from active learning should be considered for example selection,
global information can be used for bulk labeling and understanding model limitations,
while local visualizations \cite{Das2020TaxonomyMethod} can be useful for guiding granular teaching tasks.
To build modular and flexible MT systems, we recommend using model-agnostic interpretations, i.e. interpretability methods that are not specific to a particular set of models or architectures, in machine state as far as possible to minimize constraints on ML algorithms employed by the \machinelearner. 

\textit{How can we find the most effective presentation for some content?}
An effective presentation of the machine state should be able to construct the right mental model in teacher's mind about the machine.
For achieving this, we can list out common representations or visualizations of the content and conduct user evaluations to determine the most effective strategy. 
The presentation may contain interactive graphical representations  \cite{Settles2011ClosingInstances,Ramos2020InteractiveModels}, data visualizations \cite{Simard2014ICE:Problems,Taneja2022ATeaching} or text-based communication \cite{He2016Human-in-the-LoopParsing} depending on the domain and modality of the content.
In \cite{Eiband2018BringingPractice}, authors worked on improving transparency and explainability of an intelligent fitness coach using iterative prototyping and participatory design. 
Their process was guided by a comparison between the user mental model elicited by the interface about the machine with the target mental model.  
By conducting surveys and semi-structured interviews with teachers, we can determine how well the they can understand the target content presented on the teaching interface.

\subsection{\UXUI~Design in MT Systems}
\label{subsec:UI_UX}

\begin{figure*}[ht!]
    \centering
    \includegraphics[width=0.95\linewidth]{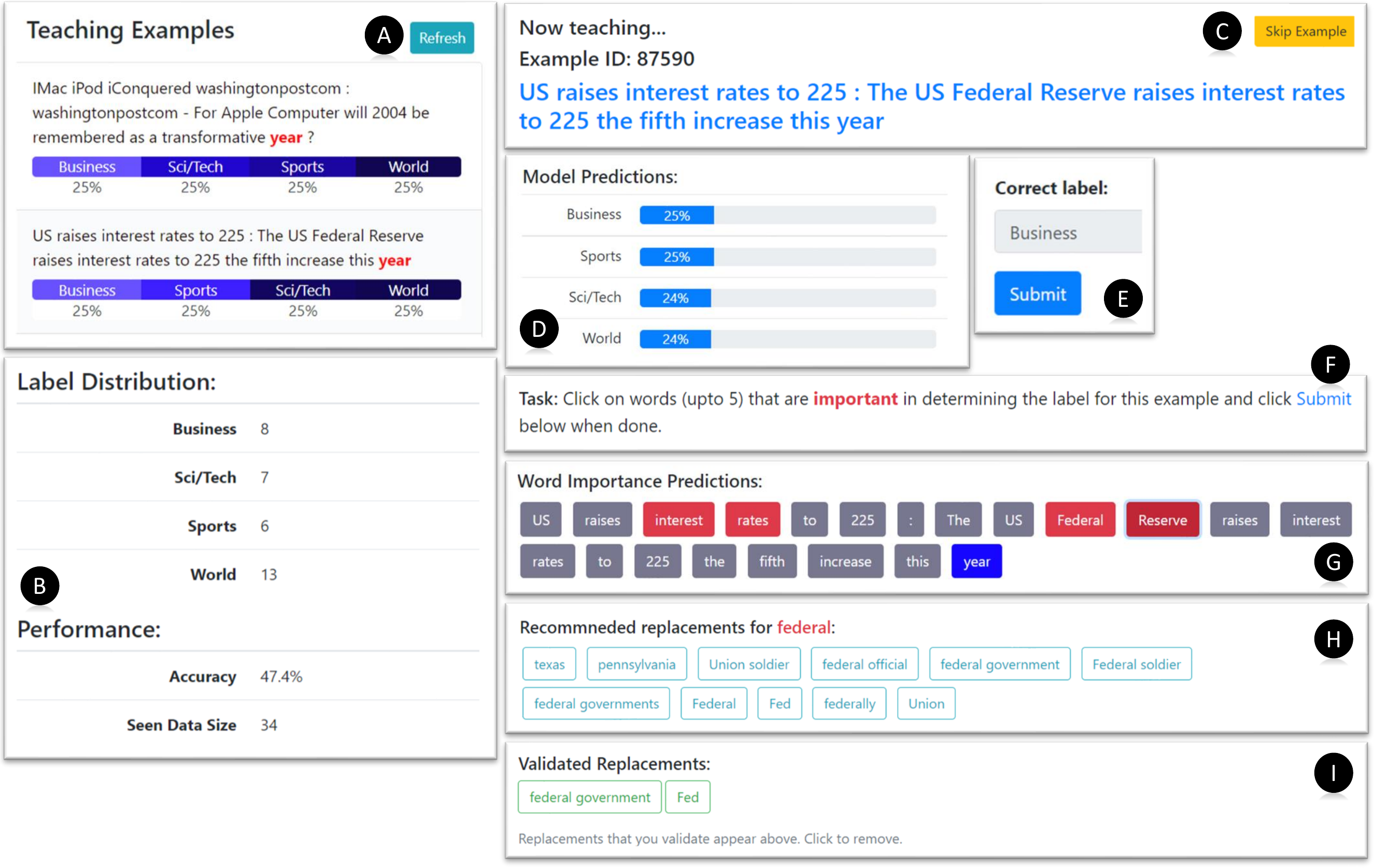}
    \caption{
    \textbf{Main elements of the web-based \teachinginterface~for our MT system are shown in panels A to I.} 
    MT system is being used here for the AG News Classification task.  
    Panel A shows confusing examples displayed to teacher for selection, along with current model prediction and potentially important words highlighted in red.
    B displays labeled data statistics and accuracy on a test set.
    C shows the example selected by teacher for reference.
    D displays current models predictions and teacher can click the right one to submit in E. 
    F describes the teaching task for highlighting important words. 
    In G, the words deemed important by the machine are highlighted in blue; marking a word by clicking it highlights it in red.
    H shows recommended replacements for words marked as important by the teacher. 
    I shows replacements that are validated by the teacher.
    }
    \label{fig:web_interface}
\end{figure*}

\textit{Is there a \UI~for the system proposed in \cite{Taneja2022ATeaching}?}
Yes, after our experiments with a command-line interface in previous work, we developed a web interface for conducting the user experiments. 
Main elements of this interface are shown in Figure \ref{fig:web_interface}. 
This frontend is supported by a
Django\footnote{\href{https://docs.djangoproject.com/en/4.0/}{https://docs.djangoproject.com/en/4.0/}} 
backend along with our existing MT system to expose the \teachinginterface~over the web.  
As shown in Figure \ref{fig:web_interface}, we have used the MT system for the AG News  Classification task \cite{Gulli2005AGsArticles}.
The user experiments are a work-in-progress but we will qualitatively discuss insights from our development process and the pilot study in this paper.

\textit{How do we get started on the \UX~design for an MT system?} 
While a detailed discussion on \UX~design is outside the scope of this paper, 
we wish to highlight two considerations for the interaction flow design, 
in particular about the order of teaching tasks. 
First, the higher level or abstract tasks may naturally come before tasks that involve more detailed analysis by the teacher.
For instance, in the interface shown in Figure \ref{fig:web_interface}, classification (panel E) is a higher level task than choosing important or inconsequential words (panel F) which is a higher level task than suggesting word replacements (panel I) that requires the most thoughtful reasoning.
Second, it may be prudent to combine or group together teaching tasks when they involve interpreting the same machine state to reduce the teaching cost.
Again, for example, in the interface shown in Figure \ref{fig:web_interface}, both important and inconsequential words tasks require interpreting the word importance values (see panel G) and are, therefore, presented to human teachers one after the other.

\textit{Similarly, are there any considerations for the \UI~design?}
Yes. 
\UI~design can play an important role in determining teaching efficiency. 
Intuitive interfaces build the right expectations in teacher's mind in terms of the functionality of interactive elements present in the interface.  
Teachers should neither be overloaded with information through the machine state, nor be under-informed to be able to provide useful feedback.
Another trade-off to consider is between the flexibility and the complexity of the teaching task. 
By increasing flexibility or decreasing constraints, the perceived complexity of the task increases because of greater number of options available to the teacher.
By having more constraints on feedback inputs, the task may get easier but teacher may not be able to thoroughly express their knowledge.
In the interface presented in Figure \ref{fig:web_interface}, teachers are not able to suggest replacements for phrases, but only words. 
This limits their ability to express novel replacements for phrases, such as replacing `New York Stock Exchange' with `Wall Street', but this also reduces the complexity of the task for teachers.

\subsection{Teaching Teachers to Teach}
\label{subsec:onboarding}

\textit{What does `teaching teachers to teach' mean?} 
Teaching a machine requires task-expertise as well as the ability to understand the machine state and perform the teaching tasks.
Teacher onboarding is the process of teaching teachers to teach i.e. giving them the knowledge they'll need to efficiently interact with the machine teaching system.   

\textit{Why do machine teachers needs training?}
While creators/designers of the machine teaching system understand the machine and may do a good job of teaching without the onboarding process, 
new teachers with no ML expertise need to be familiarized with the system.
The goal of the onboarding process is to bring the non-ML-experts at par with MT system experts in terms of teaching quality and efficiency.
Note that this will not require a lesson on ML or MT methods, but only a thorough walk-through of the MT interface. 

In \cite{Simard2014ICE:Problems}, authors observed that their participants with ML expertise did not show any advantage over other participants, though this paper does not specify the onboarding process. 
In a follow-up work from the same research group \cite{Ramos2020InteractiveModels}, authors note that 3-5 minutes training videos were sufficient in onboarding teachers to the level of expert teachers. 
They also observed and codified expert teaching patterns to create an assistant for teachers within their interface. 
While they did not observe any improvements in learner's performance, guided teachers were less frustrated and less exhausted compared to unguided teachers according to their survey results.

{
\textit{Is there any other precedence for the onboarding process?}
Yes, there is precedence for both structured onboarding processes and unstructured, cold start processes for teachers to get started with various \MT~systems. 
In \cite{He2016Human-in-the-LoopParsing}, experimenters gave very specific instructions and presented six examples to teachers before beginning the tasks. 
Some instructions asked teachers to overlook specific mistakes made by their system, while constraining the feedback to a list of options including `none of the above' for when there are no good choices.
As mentioned previously, \cite{Ramos2020InteractiveModels} used a short video and a virtual assistant to aid the teaching process.
On the other hand, \cite{Settles2011ClosingInstances} simply allowed teachers, recruited from their research group (all very likely to be ML experts), to interact with the system to familiarize themselves before starting the experiment. 
Since a lot of previous works did not discuss the onboarding process, we wish to encourage researchers to keep and share notes about onboarding in their future research.
}

\textit{How do we create an effective onboarding process?}
New teachers need be instructed for performing each teaching task and interpreting the associated machine state.
By enumerating over these tasks and states, we can construct a basic introduction to the MT interface which can be improved over time.
Once the development process is complete, we need to conduct pilot experiments with first-time teachers and understand their struggles in interacting with the MT system to improve the onboarding process along with the interface itself.
For our web interface, this involved giving clearer instructions, providing more examples and tips-and-tricks in the onboarding,  and adding quick help-guides within the interface.

\textit{How does the onboarding process for your web interface work?}
The onboarding process for interface shown in Figure \ref{fig:web_interface} broadly covers following topics for our new teachers who are not expected to have ML expertise:
\begin{enumerate}
    \item Machine as a student uses \textit{AI algorithms} to understand sentence patterns and learn vocabulary.
    
    \item News article classification task and description of the four classes (also shown in interface as a quick guide). 
    
    \item Selecting an example for teaching (panel A in Figure \ref{fig:web_interface}) and understanding class imbalance (panel B).
    
    \item Examining current predictions (panel D) and entering label (panel E) or skipping example (panel C) if assigning label is not possible.
    
    \item Understanding important/inconsequential words task (panel F) and machine's deemed important values (blue boxes in panel G).
    
    \item Correcting machine by giving feedback on important and inconsequential words (red boxes in panel G).
    
    \item Reviewing suggested replacements for important words (panel H) and validating replacements (panel I).
\end{enumerate}
For each teaching task, we also provide examples and suggest strategies to maximize efficiency.

\section{Future Research Directions}
\label{sec:future_research}

\textit{What interesting research directions emerge from this work?} %
We talk about five future research directions below to understand human-AI interactions in MT systems from the perspectives of Human-Computer Interaction (HCI) and Cognitive Science (CogSci). 

\begin{itemize}
    \item (HCI) \textbf{Teaching strategies may depend on teaching interface:} 
    Depending on the reported machine state, teacher may choose different teaching strategies. 
    For example, displaying error rate metrics may affect how teacher interacts with machine over time, or, teacher may spend less time on an example when its confusion score is low.
    An interesting research study would be to investigate how different machine states can affect human teaching and human performance.

    \item (HCI) \textbf{MT systems may improve quality of data:} 
    Typical ML workflow involves human annotator labeling one example after the other to maximize the size of training data.
    MT systems involve deeper understanding of the example for human teachers which may allow them to realize their mistakes more often as they provide additional feedback. 
    It will be interesting to study if this leads to better human performance and quality of training data.

    \item (HCI) \textbf{Design process may show diminishing returns in teaching efficiency:} 
    The process of perfecting an MT interface is both time-consuming and costly because it requires many iterations and expensive user studies.
    Therefore, it is important to study how the law of diminishing returns holds on the investment in interface design with respect to gains in teaching efficiency.

    \item (CogSci) \textbf{Mutual theory of mind in machine teaching:} 
    If we can understand how machine states relate to teaching behaviors, we can imagine building a mental model of human teacher in the machine learner building a mutual theory of mind. 
    This mental model of teacher can be useful in adapting machine behavior to draw out maximum efficiency from the human teacher.
    \cite{Wang2021TowardsAssistant} explore similar idea in a setting where a machine is the teacher, and humans are students.
    
    \item (CogSci) \textbf{Cognitive load may be lower for MT systems:} 
    In a typical annotation process, where only labels are provided by human annotators, cognitive load may be higher than in using MT systems because of higher context switching. 
    A context switch happens when the teacher starts teaching a new example from scratch. 
    We wish to study impact of MT systems on teacher engagement and their performance as a result.

\end{itemize}

\section{Summary}

An MT system has many components that work together to collect feedback from human teachers and use it to train an ML model.
A design process for MT system, and the \teachinginterface~in particular, is essential to ensure that teaching efficiency and learner's performance goals of an MT system are met.   
The \teachinginterface~design process involves creating easy and unambiguous teaching tasks, building the right mental model in teacher's mind about the learner through machine state, designing \UX~and \UI~for teachers to express their expertise, and creating an onboarding process for the new teachers. 
This paper studied design considerations and examples for each of these steps and discussed future research directions for human-AI interaction in \MT~systems.

\bibliographystyle{named}
\bibliography{ijcai22-multiauthor}

\begin{thebibliography}{}

\bibitem[\protect\citeauthoryear{Cui \bgroup \em et al.\egroup
  }{2021}]{Cui2021UnderstandingLearning}
Y~Cui, P~Koppol, H~Admoni, et~al.
\newblock {Understanding the Relationship between Interactions and Outcomes in
  Human-in-the-Loop Machine Learning}.
\newblock In {\em IJCAI}, pages 4382--4391, 2021.

\bibitem[\protect\citeauthoryear{Das \bgroup \em et al.\egroup
  }{2020}]{Das2020TaxonomyMethod}
S~Das, N~Agarwal, D~Venugopal, et~al.
\newblock {Taxonomy and Survey of Interpretable Machine Learning Method}.
\newblock In {\em IEEE SSCI}, pages 670--677, 2020.

\bibitem[\protect\citeauthoryear{Eiband \bgroup \em et al.\egroup
  }{2018}]{Eiband2018BringingPractice}
M~Eiband, H~Schneider, M~Bilandzic, et~al.
\newblock {Bringing Transparency Design into Practice}.
\newblock In {\em ACM IUI}, 2018.

\bibitem[\protect\citeauthoryear{Godbole \bgroup \em et al.\egroup
  }{2004}]{Godbole2004DocumentLabels}
S~Godbole, A~Harpale, S~Sarawagi, et~al.
\newblock {Document classification through interactive supervision of document
  and term labels}.
\newblock In {\em PKDD}, 2004.

\bibitem[\protect\citeauthoryear{Goel and
  Polepeddi}{2019}]{Goel2019JillEducation}
A~Goel and L~Polepeddi.
\newblock {Jill Watson: A Virtual Teaching Assistant for Online Education}.
\newblock In {\em Education at scale: Engineering online teaching and learning.
  NY: Routledge}. Georgia Institute of Technology, 2019.

\bibitem[\protect\citeauthoryear{Gulli}{2005}]{Gulli2005AGsArticles}
A~Gulli.
\newblock {AG's corpus of news articles}, 2005.

\bibitem[\protect\citeauthoryear{Harpstead \bgroup \em et al.\egroup
  }{2018}]{Harpstead2018TowardsFramework}
E~Harpstead, C.~J Maclellan, R.~P Marinier~Iii, et~al.
\newblock {Towards Natural Cognitive System Training Interactions: A
  Preliminary Framework}.
\newblock 2018.

\bibitem[\protect\citeauthoryear{He \bgroup \em et al.\egroup
  }{2016}]{He2016Human-in-the-LoopParsing}
L~He, J~Michael, M~Lewis, et~al.
\newblock {Human-in-the-Loop Parsing}.
\newblock In {\em EMNLP}, pages 2337--2342, 2016.

\bibitem[\protect\citeauthoryear{Jandot \bgroup \em et al.\egroup
  }{2016}]{Jandot2016InteractiveClassification}
C~Jandot, P~Simard, M~Chickering, et~al.
\newblock {Interactive Semantic Featuring for Text Classification}.
\newblock {\em arXiv:1606.07545}, 2016.

\bibitem[\protect\citeauthoryear{Kreutzer \bgroup \em et al.\egroup
  }{2018}]{Kreutzer2018CanFeedback}
J~Kreutzer, S~Khadivi, E~Matusov, et~al.
\newblock {Can Neural Machine Translation be Improved with User Feedback?}
\newblock In {\em NAACL HLT}, pages 92--105, 2018.

\bibitem[\protect\citeauthoryear{Liu \bgroup \em et al.\egroup
  }{2017}]{Liu2017IterativeTeaching}
W~Liu, B~Dai, A~Humayun, et~al.
\newblock {Iterative Machine Teaching}.
\newblock 2017.

\bibitem[\protect\citeauthoryear{Qiao \bgroup \em et al.\egroup
  }{2017}]{Qiao2017ExploringAnswering}
T~Qiao, J~Dong, and D~Xu.
\newblock {Exploring Human-like Attention Supervision in Visual Question
  Answering}.
\newblock In {\em AAAI 2018}, 9 2017.

\bibitem[\protect\citeauthoryear{Ramos \bgroup \em et al.\egroup
  }{2020}]{Ramos2020InteractiveModels}
G~Ramos, C~Meek, P~Simard, et~al.
\newblock {Interactive Machine Teaching: A Human-centered Approach to Building
  Machine-learned Models}.
\newblock {\em Human–Computer Interaction}, 35(5-6):413--451, 2020.

\bibitem[\protect\citeauthoryear{Settles}{2011}]{Settles2011ClosingInstances}
B~Settles.
\newblock {Closing the Loop: Fast, Interactive Semi-Supervised Annotation With
  Queries on Features and Instances}.
\newblock In {\em EMNLP}, pages 1467--1478, 2011.

\bibitem[\protect\citeauthoryear{Simard \bgroup \em et al.\egroup
  }{2014}]{Simard2014ICE:Problems}
P~Simard, D~Chickering, A~Lakshmiratan, et~al.
\newblock {ICE: Enabling Non-Experts to Build Models Interactively for
  Large-Scale Lopsided Problems}.
\newblock {\em arXiv:1409.4814}, 2014.

\bibitem[\protect\citeauthoryear{Simard \bgroup \em et al.\egroup
  }{2017}]{Simard2017MachineSystems}
P~Simard, S~Amershi, D~Chickering, et~al.
\newblock {Machine Teaching: A New Paradigm for Building Machine Learning
  Systems}.
\newblock {\em arXiv: 1707.06742}, 2017.

\bibitem[\protect\citeauthoryear{Stiennon \bgroup \em et al.\egroup
  }{2020}]{Stiennon2020LearningFeedback}
N~Stiennon, L~Ouyang, J~Wu, et~al.
\newblock {Learning to summarize from human feedback}.
\newblock In {\em NeurIPS}, volume~34, 2020.

\bibitem[\protect\citeauthoryear{Taneja \bgroup \em et al.\egroup
  }{2022}]{Taneja2022ATeaching}
K~Taneja, H~Sikka, and A~Goel.
\newblock {A Framework for Interactive Knowledge-Aided Machine Teaching}.
\newblock {\em arXiv:2204.10357}, 2022.

\bibitem[\protect\citeauthoryear{Wang \bgroup \em et al.\egroup
  }{2021}]{Wang2021TowardsAssistant}
Q~Wang, K~Saha, E~Gregori, et~al.
\newblock {Towards Mutual Theory of Mind in Human-AI Interaction: How Language
  Reflects What Students Perceive About a Virtual Teaching Assistant}.
\newblock 15, 2021.

\bibitem[\protect\citeauthoryear{Zhu \bgroup \em et al.\egroup
  }{2018}]{Zhu2018AnTeaching}
X~Zhu, A~Singla, S~Zilles, et~al.
\newblock {An Overview of Machine Teaching}.
\newblock 2018.

\bibitem[\protect\citeauthoryear{Zhu}{2015}]{Zhu2015MachineEducation}
X~Zhu.
\newblock {Machine Teaching: An Inverse Problem to Machine Learning and an
  Approach Toward Optimal Education}.
\newblock In {\em AAAI}, volume 29(1), 2015.

\end{thebibliography}

\end{document}